\documentclass[letterpaper,twocolumn,fleqn]{article} 

\usepackage{ist}
\usepackage{nameref}
\usepackage{amsmath}
\usepackage{amssymb}
\usepackage{booktabs}
\usepackage{tikz,url}
\usepackage{subcaption}

\title{Image Denoising with Control over Deep Network Hallucination}

\author{Qiyuan Liang, Florian Cassayre, Haley Owsianko, Majed El Helou, Sabine S\"usstrunk\\
School of Computer and Communication Sciences, EPFL, Lausanne, Switzerland.}

\date{} 

\begin{document} 

\newcommand{\medits}[1]{\textcolor{red}{#1}}
\newcommand{\figcap}[0]{\vspace{0.2cm}}
\maketitle 

\thispagestyle{empty} 


\begin{abstract}
Deep image denoisers achieve state-of-the-art results but with a hidden cost. As witnessed in recent literature, these deep networks are capable of overfitting their training distributions, causing inaccurate hallucinations to be added to the output and generalizing poorly to varying data. For better control and interpretability over a deep denoiser, we propose a novel framework exploiting a denoising network. We call it controllable confidence-based image denoising (CCID). In this framework, we exploit the outputs of a deep denoising network alongside an image convolved with a reliable filter. Such a filter can be a simple convolution kernel which does not risk adding hallucinated information. We propose to fuse the two components with a frequency-domain approach that takes into account the reliability of the deep network outputs. With our framework, the user can control the fusion of the two components in the frequency domain. We also provide a user-friendly map estimating spatially the confidence in the output that potentially contains network hallucination.
Results show that our CCID not only provides more interpretability and control, but can even outperform both the quantitative performance of the deep denoiser and that of the reliable filter, especially when the test data diverge from the training data. 


\end{abstract}

\section{Introduction}
\let\thefootnote\relax\footnotetext{Our code and models are publicly available at\\ \url{https://github.com/IVRL/CCID}}
Additive Gaussian image denoising is one of the most fundamental tasks in image restoration due to its numerous theoretical and practical uses. Various denoising problems can be reduced to Additive White Gaussian Noise (AWGN) removal with variance stabilization transforms, and AWGN denoisers can also serve as regularizers~\cite{reehorst2018regularization,el_2021}.  Although denoisers are omnipresent in imaging pipelines, the denoising problem remains inherently challenging due to its ill-posed nature, and the difficulty of image quality assessment itself~\cite{quality_assessment}.

Classic denoisers proposed various image priors, adding constraints to reduce the ill-posed nature of the image denoising problem~\cite{bm3d,WNNM}. With the emergence of deep learning, novel solutions were developed, outperforming the classic methods both quantitatively and qualitatively~\cite{dncnn,mwcnn,noise2noise}. However, deep learning methods have their own drawbacks, such as poor generalization on unseen images from another domain and images with different noise levels~\cite{SFM}. Deep networks are also inherently a black box, making it difficult to interpret their outputs. When deep networks make mistakes, having control over the system and the ability to reason about the output becomes  necessary~\cite{confidence_classification_1}.

\begin{figure}[t]
\begin{center}
\includegraphics[width=\linewidth]{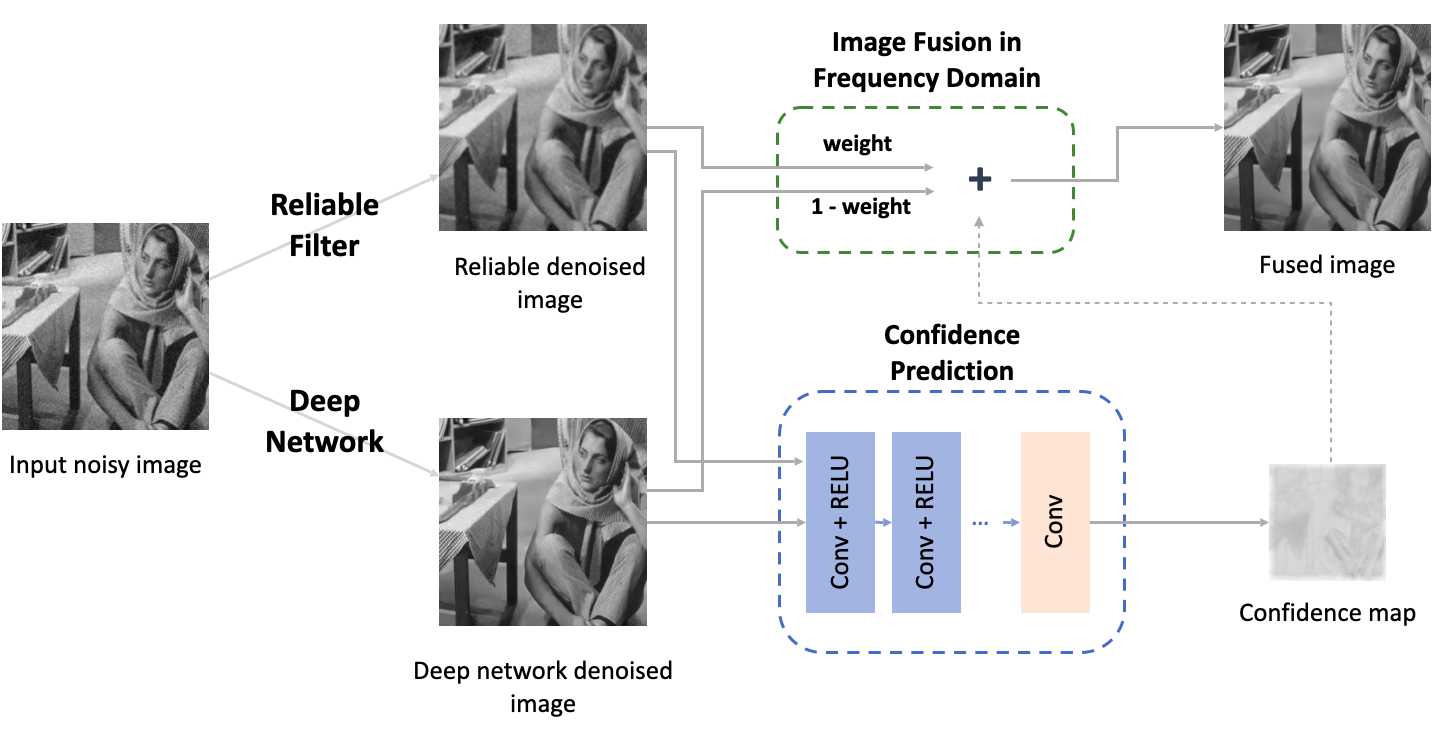}
\end{center}
\caption{Our controllable confidence-based image denoising (CCID). The input image is denoised in parallel with a convolution filter and with a deep neural network, and the final output is obtained by fusing the two results. The deep network output is further fed into the confidence prediction network to obtain a confidence map. Users can visualize this map to interpret the network's result, or additionally apply it to steer the overall fusion.}
\label{fig:pipeline}
\end{figure}

To address the aforementioned problems, multiple approaches were proposed, including increasing the overall robustness~\cite{brownlee2018better}, modeling the confidence value~\cite{confidence_classification_1}, detecting out-of-distribution (OOD) data~\cite{confidence_classification_2}, or improving generalization~\cite{SFM, denoising_overfitting}. Here, we present a controllable confidence-based image denoising (CCID) method. It is a novel framework addressing generalization and interpretability in image denoising by exploiting reliable convolution filters and giving control and insight to the user.
As shown in Figure~\ref{fig:pipeline}, CCID denoises images with a convolution filter in parallel to the deep network. Users fuse the two outputs to produce the final denoised image by exploiting the image denoised by convolution, the deep network denoised image, and a predicted confidence map. 

Our contributions can be summarized as follows. We propose and evaluate frequency-domain methods to fuse an image denoised by convolution and the deep network denoised image. 
We provide a confidence prediction that reveals the regions where the deep denoiser is likely to produce an error. We thus give users the flexibility to smoothly fuse the image denoised by convolution and the deep denoised image in the frequency domain based on confidence predictions. This framework enables users to exploit the convolution denoising when the deep network is likely inaccurate, hence safeguarding against incorrect network hallucinated information. Our experimental results show that beyond providing this control to users, our CCID outperforms both of its underlying denoisers quantitatively. Such is the case with data from a different domain, noise level, or noise type than that of the training data.


\section{Related Work}\label{literature_review}
\textbf{Image denoising} is a widely studied problem in the literature, particularly the fundamental AWGN removal. Classic image denoising methods make various assumptions to improve denoising. These assumptions form a prior that can improve over the maximum likelihood solution for this ill-posed problem~\cite{blind_bayesian}. For instance, BM3D makes the assumption that there are similar patches within an image~\cite{bm3d} and applies collaborative filters to groups of similar patches. WNNM assumes that images can be represented as low-rank matrices~\cite{WNNM} and exploits this attribute to filter out noise components.

With the advent of deep learning methods, richer priors could be learned over training image distributions. Deep learning thus achieved superior denoising performance, notably with DnCNN~\cite{dncnn}, where a residual connection is used to directly predict the noise map rather than the denoised image, thus alleviating the learning burden~\cite{dncnn}. MWCNN introduces a new form of downsampling and upsampling layers for building deep networks in the discrete wavelet domain~\cite{mwcnn}. Feeding the network with differentiated low- and high-frequency data guides the learning process and hence improves the performance. DDFN integrates a dilated convolutional network with the deep boosting framework to further boost the denoising performance~\cite{deep_boosting}. To deal with blind noise settings, FFDNet employs a customizable noise level map as input~\cite{FFDNet}, which, although not practical in test settings, can improve the results. BUIFD presents a blind and universal denoiser for AWGN removal~\cite{blind_bayesian} improving the generalization strength. This is achieved by exploiting the maximum a priori solution under a theoretical prior and integrating it into the network's architecture along with an internal noise map estimation. 

Although deep networks significantly improve the denoising results, they do so by hallucinating data based on their rich learned priors. This reliance on learned priors can cause inaccurate hallucinations, particularly when the test image lies outside the training distribution. What we call \textit{hallucination} is the information derived from prior information and not directly inferred from observed data. Unlike previous methods, we propose a framework to wrap deep denoisers with reliable filtering techniques, and thus provide control over the network hallucination.

\textbf{Confidence in deep learning} is a critical problem. While neural networks achieve high accuracy, they can also provide unreliable outputs~\cite{confidence_classification_0}. The issue is that these unreliable outputs can have high confidence. Due to the black-box nature of such networks, the lack of interpretability does not enable any safe-guarding against such errors that can be critical in certain applications like surveillance~\cite{aakerberg2021real}. This problem is aggravated in the case of test data that are Out Of Distribution (OOD) relative to the training set. To address this issue, a baseline for assessing network confidence was introduced in classification. It is based on the insight that successfully classified examples have higher maximum softmax probabilities than incorrectly classified OOD examples~\cite{confidence_classification_2,confidence_classification_3}.

Confidence has thus been explored in classification~\cite{lin2021fidelity}, but in the field of image restoration, it has received less attention. Two recent papers developed an architecture that incorporates confidence within the neural network to guide the deraining process, with the confidence map calculated at multiple levels to obtain higher accuracy~\cite{uncertainty_deraining_0,uncertainty_deraining_1}. The epistemic uncertainty of deep denoisers is exploited in~\cite{ma2021deep} to create a virtual self ensemble and improve denoising results. In contrast, our method estimates confidence to provide more control and interpretability to the user of a denoising method. This confidence can be exploited to guide the fusion of the reliable denoising outputs and the deep denoiser outputs.
The closest work to ours is the BIGPrior framework~\cite{bigprior}. It decouples the deep network hallucination from data fidelity by exploiting a pretrained generative network projection prior. This framework provides more interpretability to the user over deep hallucination. In contrast, we additionally provide control to the user. We enable the user to control the contribution of deep network outputs with a carefully designed frequency fusion, while providing an estimated confidence map.

\textbf{Image fusion} is the process of combining complementary images from multiple sensors, temporal domains, viewpoints, etc.~\cite{fusion_survey}. We focus on the fusion of the same image but restored using complementary methods. Classic image fusion techniques can be divided into those carried out in the spatial or in the frequency domain. Direct fusion in the spatial domain can cause undesired distortions in image details~\cite{image_fusion_review}. As a result, it produces inferior results compared to fusion in the frequency domain. To work in the frequency domain, the discrete cosine transform (DCT) and the discrete wavelet transform (DWT) are often exploited by fusion algorithms~\cite{img_fusion_ref_0,img_fusion_ref_1,img_fusion_ref_2}. We make use of both transforms in our experimental evaluation.

\section{Method}\label{design}
\subsection{Denoising}
We term reliable denoisers the methods that rely on minimal priors to produce a predictable output, with practically no hallucination. These approaches, such as convolution filters, are robust against various noise settings or data distributions. This is due to their simple prior assumptions that can generalize well rather than overfit. The Gaussian filter is one such example. It acts as a low-pass filter and, as a result, causes a certain loss of sharpness but ensures no hallucinated information is added. We use the Gaussian filter in our experiments for its consistency across images, its computational efficiency, and as it requires no settings adjustments besides the choice of the kernel. Other methods such as bilateral filtering and non-local means (NLM) are also potential alternatives.

In parallel, we also exploit deep neural network denoisers in our CCID framework. Deep learning approaches outperform traditional methods, however, their rich priors are dependent on the training data and can cause incorrect hallucinations. Hence, their strength can also be their main weakness when test data differ from the training conditions (data distribution, degradation model, noise level, etc.). This can be detrimental in practical settings especially when users are not aware of the training conditions of a pretrained denoiser. In our experiments, we rely on the widely used denoising model DnCNN~\cite{dncnn}. However, any other deep denoiser can be used to replace it in CCID.

We explain in what follows the two key components in our CCID framework. They enable the fusion of both denoising techniques while providing user control over network hallucinations and interpretability.

\subsection{Confidence prediction}
We design a confidence prediction network that takes the initial noisy image, the image denoised with the convolution filter, and the noise map predicted from the deep denoising network as inputs. The objective is to determine the reliability of the noise map prediction. We observed empirically that the additional information and structural features provided by the initial noisy image and the filtered image reduce the convergence time and improve the accuracy of our confidence predictor. Our model predicts a confidence value for each $8 \times 8$ region in the image. Predicting the exact error of the deep denoising network would imply that our confidence prediction network is able to outperform the underlying denoising network by simply correcting its errors. Instead, our objective is to estimate the extent of potential error rather than the exact error itself. To that end, we define our target ground-truth confidence $c_{GT}$ to be 
\begin{equation}
    c_{GT} = 1- 
    \frac{\left\lVert y_{GT} - y_{DNN} \right\lVert_1 \downarrow_8}
    {\sigma_{max}},
\end{equation}
where $y_{GT}$ is the ground-truth noise-free image, $y_{DNN}$ is the denoised image predicted by the deep neural network, $\left\lVert \cdot \right\lVert_1$ is the $\ell_1$ norm, $\downarrow_8$ is a downsampling by a factor of $8$ carried out with an average pooling operation, and the error is normalized to $\sigma_{max}$ that we define next. In our experiments, we consider noise levels between $\sigma=0$ and $100$, as is common in the literature. Therefore, $\sigma_{max}$ is set to be $100$ for normalization, which is the highest test noise level. Using the $8 \times 8$ region provides a good balance between spatial precision and network accuracy. This region-based technique generates reasonably accurate predictions while also offering useful confidence information that enables users to identify regions that may contain incorrect hallucinations. We thus train a neural network to predict a confidence $c$ (visualized in Figure~\ref{fig:confidence_interpolation}), which is an estimate for $c_{GT}$. This network is trained with the setup provided in the experimental evaluation section.

\subsection{Proposed fusion}
We propose two modes of fusion to combine images denoised by a reliable filter and a deep denoiser. One uses a scalar parameter for the fusion, the other exploits our confidence map.
For the first type, we carry out the fusion in the frequency domain to smoothly blend the structure of both images. We considered two domain transforms, the DWT and the DCT for the fusion. The second mode builds on our DWT fusion.


\subsubsection{Wavelet and cosine fusion}
With the DWT, the fusion is computed as the weighted average of the transformed images, level by level. Two images of the same size go through the same decomposition levels of the DWT, and the coefficients at each level are retrieved accordingly. The fusion is then computed as the weighted average between the two. For DCT fusion, the spectral power across is usually modeled as a probability density function following a half-Gaussian distribution~\cite{dct_distribution}. We propose a mask-based fusion approach in the DCT domain based on this observation and empirical assessments. The mask is given by
\begin{align}
& M_w(\omega_x,\omega_y) = \exp(-(\omega_x^2+\omega_y^2)/(2s)), \nonumber\\
& \text{where } s = a\left(\frac{1}{1-w+\epsilon}-1\right),
\end{align}
and where $\omega_x$ and $\omega_y$ are the DCT frequency coefficient indices; $w$ is the user-controlled fusion weight; $a$ is a scale factor ($0.1$ in practice); and $\epsilon$ is a small positive offset ($1\mathrm{e}{-3}$ in practice) to maintain the numerical stability.

The mask is used to perform the interpolation between the values of each frequency, with the output matching the deep network denoised image when $M=1$ and the reliable image when $M=0$. The low frequencies of the deep denoised image are fused first, followed by higher frequencies as the fusion weight increases. This is because low frequencies are easier to reliably predict for most image restoration tasks, hence we integrate them into our output before higher frequencies.

\subsubsection{Confidence-aware fusion}
Our second, confidence-aware, fusion builds on the previous functions but additionally relies on the confidence map. The latter contains spatial reliability information that enables us to perform a fine-grained fusion. The goal is to preserve the details when the risk of error is low, and otherwise average them with the reliable filter to mitigate that risk. Note that the confidence map is only an estimation and can itself be prone to errors. However, as demonstrated in our experimental evaluation, the addition of this component is still beneficial (Figure~\ref{fig:psnr_ssim_conf}). With the DWT, we introduce a patch-wise fusion, applying the DWT on each $8 \times 8$ region and stitching back the fusion results. Each $8 \times 8$ region has a confidence value, therefore, we use the confidence value per region to guide the fusion. We modulate the fusion weight on each region such that
\begin{equation}\label{eqn:updated_conf}
    w_{region} = w \cdot (1 + c - t),
\end{equation}
where $w$ is again the user-controlled fusion weight, $c$ is the region's confidence value, and $t$ is a fixed threshold. If the confidence value is higher than the threshold, we give the deep denoised image more weight to obtain a better denoising result. The threshold is set to $0.8$ in practice based on the in-distribution and OOD data, as discussed in the following section.




\section{Experimental Evaluation}\label{results}
We use DnCNN~\cite{dncnn} (trained on Gaussian noise $\sigma=25$) as the deep network and a Gaussian kernel as the reliable denoising filter in our experiments. As mentioned earlier, other methods can replace these denoisers in a straight-forward manner in our CCID. 

\subsection{Controllable fusion}
\begin{figure}[t]
\captionsetup[subfigure]{labelformat=empty}
\begin{center}
\input{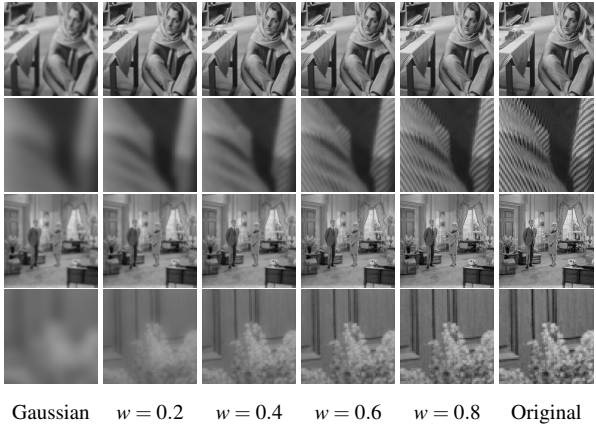}
\end{center}
\figcap{}
\caption{Visualization of our fusion techniques with different fusion weights $w$. The Gaussian filter is a kernel with $\sigma=4$. The deep denoised image is replaced by the ground-truth for comparison purposes. The first two rows show the DCT fusion results, and the last two show the DWT ones (with zoom-in crops). Observe that the edges gradually become sharper with increasing $w$.\\}
\label{fig:controllable_fusion}
\end{figure}

\begin{figure}[t]
\begin{center}
\scalebox{0.32}{\input{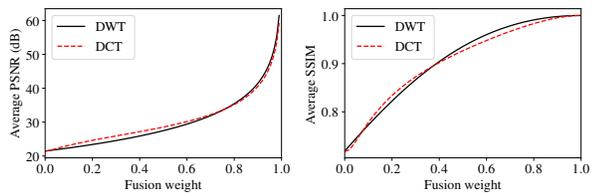}}
\end{center}
\caption{Quantitative evaluation of the DCT and DWT fusion on the BSD68~\cite{bsd} image dataset. The experimental settings are identical to those in Figure~\ref{fig:controllable_fusion}. Both approaches provide similar fusion performance, with an advantage to DWT on larger weights.}
\label{fig:dct_quantitative}
\end{figure}

Figure~\ref{fig:controllable_fusion} shows the qualitative results of the proposed fusions, with DWT and with DCT. We replace the deep denoised images with clean images for illustrative purposes, and the Gaussian filter is applied to the clean images. As can be observed, the fusion progresses smoothly from the Gaussian-filtered to the clean image, with the increasing $w$ weight. In the zoom-in crops, we see the edges and textures monotonically appearing in the second and fourth row of Figure~\ref{fig:controllable_fusion}. 
We also evaluate the results using quantitative measurements, with the peak-signal-to-noise ratio (PSNR) and the structural similarity index measure (SSIM). The corresponding results are shown in Figure~\ref{fig:dct_quantitative}. The progressive increase in PSNR and SSIM is in accordance with our qualitative observations, with added high frequencies as the weight increases. For the given metrics, the DWT and DCT fusion algorithms produce relatively similar results. This smooth fusion in the frequency domain enables the CCID users to control the contribution of reliable denoising and deep denoising in their final output.

\subsection{Confidence prediction}
\begin{figure}[t]
\centering
\input{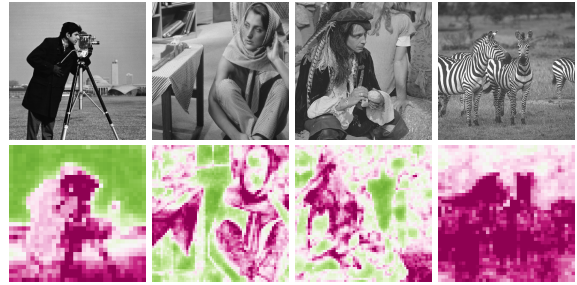}
\figcap{}
\caption{Visualization of the confidence maps for noisy images ($\sigma=25$). The threshold is $0.95$ for better visualization, and all the values above it are green, while those below it are purple. The color intensity represents the distance to the threshold. The confidence map is $8$ times smaller than the input stack.\\}
\label{fig:confidence_interpolation}
\end{figure}
The training data of our confidence prediction network comprise varying noise levels. It is trained on the same data generated by the DnCNN network trained for Gaussian noise removal. The data could potentially be further extended to cover different noise types and image domains. Our network produces meaningful and precise confidence maps on held out test sets. Figure~\ref{fig:confidence_interpolation} shows the predicted confidence map for each corresponding image. The confidence value is high for low-frequency regions, such as the sky, walls, and plain T-shirts. The confidence value is relatively low for high-frequency regions, such as meadows, table cloth, and hair. Indeed, high-frequency component corruption is harder to correct. As a result, high-frequency regions are areas where deep networks are most likely to produce hallucinated errors~\cite{SFM}, resulting in a generally lower confidence score.

\subsection{Fusion with confidence}
\begin{figure}[t]
\begin{center}
\scalebox{0.32}{\input{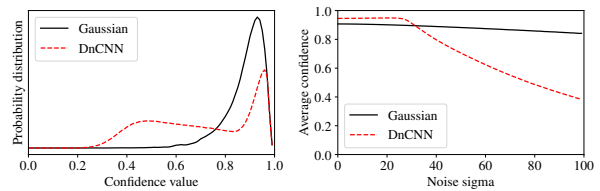}}
\end{center}
\caption{Confidence distribution (left) for the Gaussian filter and DnCNN, for noise levels $\sigma=0$ to $100$. Average confidence value (right) as a function of the noise level. DnCNN confidence values decrease for increasing noise levels after 0.4-0.5, which is corresponding to its training noise level. As the noise sigma increases, the average confidence of the Gaussian filter drops slowly, while that of DnCNN drops quickly after sigma exceeds 25.}
\label{fig:conf_dist}
\end{figure}

\begin{figure}[t]
\captionsetup[subfigure]{labelformat=empty}
\begin{center}
\input{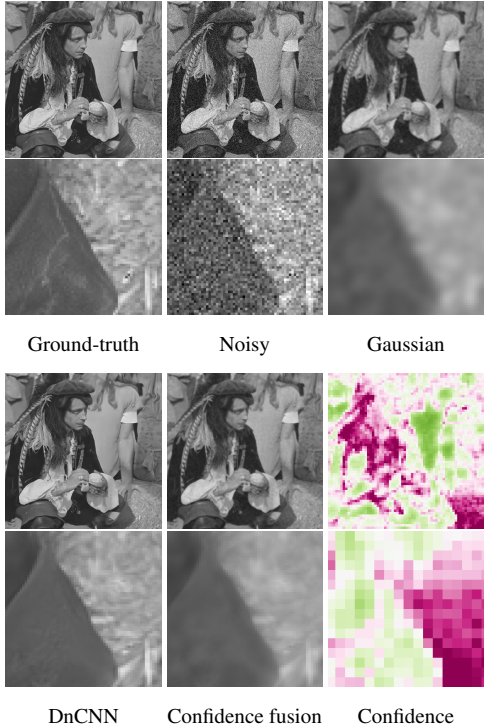}
\end{center}
\figcap{}
\caption{Confidence-aware DWT fusion for noise level $\sigma=25$. On the zoomed-in images (bottom row), the left part has higher confidence than the right part of the crop. The fused output is both sharp and reliable for the left part. The right part is mostly based on the reliable-denoiser output and is thus blurry. Note that we aim to be safe rather than overconfident, which pays off in terms of quantitative performance for out-of-distribution data.}
\label{fig:conf_guided_fusion}
\end{figure}

\begin{figure}[t]
\begin{center}
\scalebox{0.32}{\input{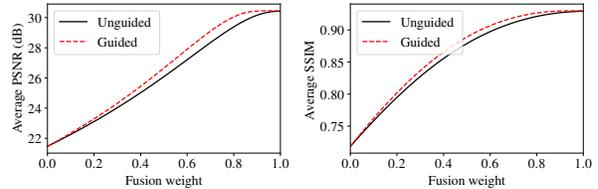}}
\end{center}
\caption{Quantitative evaluation of the unguided and confidence-guided fusion on BSD68~\cite{bsd} with varying fusion weight.\\}
\label{fig:psnr_ssim_conf}
\end{figure}

\begin{figure}[t]
\captionsetup[subfigure]{labelformat=empty}
\begin{center}
\input{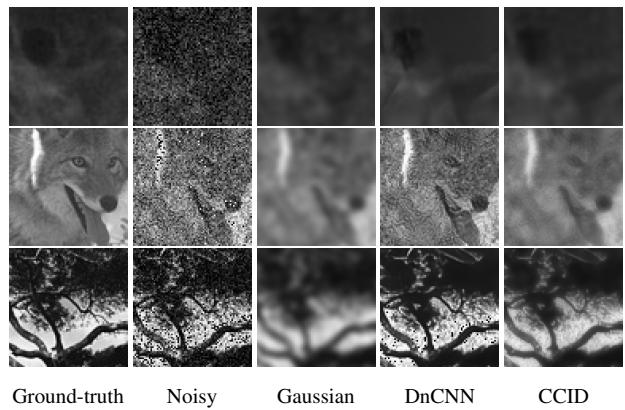}
\end{center}
\figcap{}
\caption{Sample zoomed-in fusion outputs that optimize PSNR and SSIM on OOD data. The corresponding fusion weights are $0.25$, $0.2$, and $0.4$. Row 1 is from the FMD dataset~\cite{fmd_dataset} with noise level $25$. Row 2 comes from the BSD dataset with Gaussian noise $\sigma=35$. Row 3 is from the BSD dataset with Poisson noise. In our DWT fused output, the incorrect hallucinations introduced by DnCNN are mitigated, at the cost of sharpness.}
\label{fig:ood_fusion}
\end{figure}

To determine the appropriate threshold for Equation~\eqref{eqn:updated_conf}, the results of the reliable filter and DnCNN need to be differentiated. We plot the confidence value distribution for the two methods in Figure~\ref{fig:conf_dist}. When varying the noise level from $\sigma=0$ to $100$ in our experiments, the results for the Gaussian filter have an exponential distribution, whereas DnCNN's distribution is bi-modal. DnCNN is only trained on noise level $\sigma=25$, and does not generalize very well. The peak of confidence values around 0.4-0.5 are due to its better-than-average performance on noise level 25. For the other noise levels, the confidence drops on average with increasing noise levels for both denoisers, albeit faster for the deep network. To differentiate the two approaches, we fix our threshold at $0.8$ in Equation~\eqref{eqn:updated_conf} for our experiments.

Figure~\ref{fig:conf_guided_fusion} shows the fusion results using a confidence guide. The zoomed-in regions show two areas with disparate confidence values; the simple fabric on the left has a higher confidence value than the threshold, while the right textured part has a lower confidence value. For the confidence-aware fusion, we intentionally put a higher weight on the confident region by scaling the global fusion weight using Equation~\eqref{eqn:updated_conf}. The confidence-aware method performs slightly better than the fusion method without confidence by preserving the confident component from the deep denoiser. The proposed guide does not significantly improve the overall quantitative performance for the in-distribution denoising results, as shown in Figure~\ref{fig:psnr_ssim_conf}. However, it gives more control and interpretability to the users. When the test set is similar to the training set, we almost always achieve the best performance at $w=1$ (deep denoised image). However, this is not the case with OOD data. In real-life scenarios, it is common to deal with data that is not similar to the training set. We study this configuration in the next section.

\subsection{Out-of-distribution data}
\label{OOD}
We consider three types of OOD data, namely, images from another domain, images with noise levels different from the ones in the training set, and images with a different noise type. As shown in Figure~\ref{fig:ood_fusion} and Table~\ref{tab:ood_fusion}, the fusion no longer achieves the best performance with $w=1$. Instead, in certain cases, the results obtained with a Gaussian filter are already better than those of DnCNN. The DnCNN network achieved a better performance than the Gaussian filter only on the dataset from another domain, but not for varying noise. 
These results highlight the generalizability problem of the deep learning network, a problem that is not faced by the Gaussian filters. For the microscopy data in Figure~\ref{fig:ood_fusion}, we can observe strange line patterns near the cell in DnCNN's results that do not exist in the original noisy image. Furthermore, as the noise level increases, DnCNN can no longer remove the noise effectively. The training set covers only the noise level $\sigma=25$, therefore, the model is not capable of predicting a noise map with a larger standard deviation. When the input is deteriorated with a different noise type, black dots appear in the denoised output of DnCNN, while the background is left noisy. Our proposed fusion algorithm employs both the reliable filter to maintain the basic structural information and the deep learning method to obtain sharper, although potentially incorrect, details. Enabling users to adjust the fusion weight offers another layer of protection against deep network hallucination. Table~\ref{tab:ood_fusion} shows that CCID outperforms both of its underlying denoisers. We show for reference the results of CCID$_d$ that uses a fixed default fusion weight, which already outperforms both denoisers on data from a different noise level or noise type. The CCID results tweaked by a user for best reconstruction outperform on all OOD types. 

%
%

\begin{table}[]
  \begin{center}
  \scalebox{0.78}{
    \begin{tabular}{l|c|c|c|c}
      \toprule
      \textbf{OOD Type} &
      \textbf{Gaussian} &
      \textbf{DnCNN} &
      \textbf{CCID$_{d}$} &
      \textbf{CCID}\\
      \hline
      Data Domain & 32.48/0.91 & 35.17/\textcolor{red}{0.95} & {34.41/0.94} & \textcolor{red}{35.20/0.95} \\
      Noise Level & 23.80/0.79 & 20.02/0.48 & 24.45/0.78 & \textcolor{red}{24.55/0.83}\\
      Noise Type & 23.92/0.80 & 21.60/0.62 & 24.69/0.72 & \textcolor{red}{25.01/0.81}\\
      \bottomrule
    \end{tabular}
  }
    \figcap{}
    \caption{Average PSNR and SSIM, shown in this order, on out-of-distribution (OOD) data. We vary the data domain distribution, the noise level, and the type of noise of the test images. The experimental settings are identical to those in Figure~\ref{fig:ood_fusion}. The fusion weight in CCID is tweaked by the user for best restoration. For reference, we also include for reference the results of CCID$_d$ that fixes a default invariable weight $w=0.5$.}
    \label{tab:ood_fusion}
  \end{center}
\end{table}

\section{Conclusion}\label{conclusion}
We present a novel denoising framework, CCID, where a reliable filter is exploited alongside a deep network to provide control over network hallucination. Our framework improves the overall interpretability and the generalizability of the deep network by merging it with the reliable filter. We further provide a confidence map that can either be integrated into our fusion or visualized by the user.
Our results show that our fusion provides flexibility to the user without artifacts. It also improves the robustness against out-of-distribution data, on which it outperforms both underlying denoisers. Future work can study the generalization of our framework to other restoration tasks.




\newpage
{
\small
\bibliographystyle{unsrt}
\bibliography{main}
}




\end{document}